\documentclass[twocolumn,twoside,slac_two]{revtex4}
\usepackage{graphicx}
\usepackage{fancyhdr}
\def\ifmath#1{\relax\ifmmode#1\else$#1$\fi}
\def\etal{{\it et al.}}

\def\ie{{\it i.e.}}
\def\eg{{\it e.g.}}

\def\ra     {\ifmath{\rightarrow}}

\def\babar{\mbox{\sl B\hspace{-0.4em} {\footnotesize\sl A}\hspace{-0.4em} B\hspace{-0.4em} {\footnotesize\sl A\hspace{-0.1em}R}}}

\def\epem  {\ifmath{e^+e^-}}
\def\DsJ   {\ifmath{D_{sJ}}}
\def\DsTT{\ifmath{D_{sJ}^*(2317)^+}}
\def\DsTO{\ifmath{D_s^{*}(2112)^+}}
\def\DsFE{\ifmath{D_{sJ}(2460)^+}}
\def\FourS {\ifmath{\Upsilon{\rm(4S)}}} %<<<
\def\ccbar	{\ifmath{c\bar c}}
\def\BBbar	{\ifmath{B\bar B}}
\def\gg    {\ifmath{\gamma\gamma}}

\def\piz   {\ifmath{\pi^0}}

\def\Ds    {\ifmath{D^+_s}}

\def\jpsi  {\ifmath{{J\mskip -3mu/\mskip -2mu\psi\mskip 2mu}}} %<<<
\def\pppm       {\ifmath{\pi^+\pi^-}}

\def\btoc	{\ifmath{b \ra c}}

\def\BF 	{\ifmath{{\cal B}}}

\def\rootS      {\ifmath{\sqrt s}}

\def\Bfac{$B$-Factory}
\def\Bdcy{$B$~decay}
\def\Bmeson{$B$~meson}

\def\JPC  {\ifmath{{\rm J^{PC}}}}
\def\JP   {\ifmath{{\rm J^{P}}}}

\def\ev  {\ifmath{\mbox{\,e\kern -0.08em V}}} %<<<
\def\kev  {\ifmath{\mbox{\,ke\kern -0.08em V}}} %<<<
\def\mev  {\ifmath{\mbox{\,Me\kern -0.08em V}}} %<<<
\def\gev  {\ifmath{\mbox{\,Ge\kern -0.08em V}}} %<<<
\def\gevc {\ifmath{\mbox{\,Ge\kern -0.08em V$\!/c$}}} %<<<
\def\mevc {\ifmath{\mbox{\,Me\kern -0.08em V$\!/c$}}} %<<<
\def\gevcc{\ifmath{\mbox{\,Ge\kern -0.08em V$\!/c^2$}}} %<<<
\def\mevcc{\ifmath{\mbox{\,Me\kern -0.08em V$\!/c^2$}}} %<<<

\def\invpb{\ifmath{\mbox{\,pb}^{-1}}}
\def\invfb{\ifmath{\mbox{\,fb}^{-1}}}
\def\nb   {\ifmath{\mbox{\,nb}}}
\def\pb   {\ifmath{\mbox{\,pb}}}

\begin{document}

%Title of paper
\title{Charm and Charmonium Spectroscopy at the {\boldmath $e^+e^-$} {\boldmath $B$}-Factories}

% Repeat the \author .. \affiliation  etc. as needed
%
% \affiliation command applies to all authors since the last
% \affiliation command. The \affiliation command should follow the
% other information

\author{Helmut Marsiske\footnote{Work supported by Department of Energy contract 
DE-AC02-76SF00515.}}
\affiliation{Stanford Linear Accelerator Center, 2575 Sand Hill Road, Menlo Park, CA 94025, USA}

\begin{abstract}
Over the past few years, there has been a lot of progress in the areas of charm and
charmonium spectroscopy, in large part due to the very large data samples being 
accumulated at the \epem\ B-Factories. 
In this presentation I will focus on results in three areas: 
the X/Y/Z charmonium-candidate states, the \DsJ\ charmed-strange mesons,
and newly-discovered charmed baryons.
Note the absence of a section on pentaquarks: all \Bfac\ searches for pentaquarks, charmed
or otherwise, have not yielded any observation of such states.
\end{abstract}

%\maketitle must follow title, authors, abstract
\maketitle

\thispagestyle{fancy}

% body of paper here - Use proper section commands
% References should be done using the \cite, \ref, and \label commands
% Put \label in argument of \section for cross-referencing
%\section{\label{}}

\section{Introduction}
The results I am going to present %\footnote{Due to lack of space
%in this write-up, I will only show figures that have not yet appeared in publications
%or hep-ex preprints.} 
are obtained from data collected by \babar @PEP-II, 
Belle@KEK-B, and CLEO@CESR. \babar\ and Belle data samples are of the order of
450--650 million (M) \ccbar\ and 380--550M \BBbar\ events, collected at
center-of-mass (CM) energies on or just below the \FourS\ resonance, 
$\rootS \approx 10.58\gev$.

Charm and charmonium states at the $B$-Factories are produced in 
$\epem \ra \ccbar$ annihilation events, in 
$\epem \ra \epem\, \ccbar$ two-photon events, 
and in \Bdcy\ proceeding through the dominant \btoc\ transition.
Important variations of the annihilation process are
{\it (i)} initial-state-radiation (ISR), which reduces the available CM energy of the 
\ccbar\ pair (\eg, into the region of the charmonium system), and
{\it(ii)} double-charmonium production, where an additional \ccbar\ pair 
is pulled out of the vacuum.
The annihilation process (with and without ISR) proceeds through a virtual photon, 
thus fixing the quantum numbers (QNs) of the final state to be $\JPC = 1^{--}$. 
For production via the collision of two quasi-real photons, the final state has to have
positive $C$-parity and cannot have spin 1, \ie, $\rm J^C = 0^+,\ 2^+$.
\Bdcy\ allow access to a multitude of final states with masses below the \Bmeson\ mass;
their spin-parities can be determined from analyzing appropriate decay angular distributions.
A particularly interesting \Bdcy\ is the case of a two-body decay
$\bar B \ra \bar{K}^{(*)}\,X$\footnote{Unless noted otherwise, charge conjugation
is implied throughout the text.}, where the Cabibbo-Kobayashi-Maskawa (CKM)
coupling strengths strongly favor \btoc\ and $W\ra s\bar c$ transitions,
resulting in the production of a $X=\ccbar$ system recoiling against the $\bar K^{(*)}$.

\section{Charmonium-candidate States}

Thanks in large part to the large and steadily increasing data samples at the $B$-Factories,
there has been a steady stream of discoveries of charmonium-candidate states, 
resulting in somewhat of an alphabet soup of (unimaginatively named) $X/Y/Z$ states.
Generally, there has been good progress in determining the properties of those states,
not the least because of the ability to pull together information from multiple
production mechanisms.

\subsection{$\bf X(3872)$}

The first of these states, the X(3872), was discovered by the 
Belle Collaboration~\cite{prl91.262001} in \Bdcy\ in the reaction
$B\ra K\,X(3872),\ X(3872)\ra \pppm\jpsi$, 
using a data sample of 152M \BBbar\ pairs.
Existence of the $X(3872)\ra \pppm\jpsi$ decay was quickly confirmed
by CDF, D0, and \babar.
Since then, \babar\ and Belle have analyzed samples of 232M~\cite{prd73.011101}
and 275M~\cite{hepex.0505038} \BBbar\ pairs, respectively, 
and observe relatively clean $X(3872)$ 
signals of approximately $50-60$ events.
They measure
an average product branching fraction (BF) 
\begin{eqnarray}
\BF(B\ra K\,X(3872),\ X(3872)\ra \pppm\jpsi) = \nonumber \\
(11.6 \pm 1.9) \times 10^{-6}\,,
\label{eq.1}
\end{eqnarray}
where statistical and systematic errors have been combined in quadrature,
and a mass and width
\begin{eqnarray}
m_X      & = & (3871.2 \pm 0.6)\mevcc, \nonumber \\
\Gamma_X & < & 2.3\mev\ @\ 90\%\ {\rm C.L.}
\label{eq.2}
\end{eqnarray}

The measured mass, width, and decay mode make it difficult to accommodate
the $X(3872)$ as a conventional charmonium state. Alternative interpretations
have been proposed, for example, in terms of a $\bar D^0$-$D^{*0}$ molecule~\cite{DDbarMolecule}, 
or a diquark--antidiquark state~\cite{FourQuark}.
To help establish the nature of the $X(3872)$, additional information on its
properties, like spin-parity or other decay modes, is needed---and has been 
forthcoming with the increasing \Bfac\ data samples.

\babar\ has searched inclusively~\cite{prl96.052002} 
for $B\ra K\,X(3872)$
using a \BBbar\ data sample with one $B$ fully reconstructed.
The searched-for two-body decay of the other \Bmeson\ results 
in a monochromatic line in the kaon momentum spectrum in the \Bmeson\
rest frame.
\babar\ observes signals for a number of well-known charmonium states,
with BFs consistent with exclusive measurements, but no significant
$X(3872)$ signal is found, resulting in an upper limit on $X(3872)$ 
production in charged-\Bdcy
\begin{eqnarray}
\BF(B^\pm\ra K^\pm\,X(3872)) < 3.4 \times 10^{-4}\ @\ 90\%\ {\rm C.L.}
\label{eq.3}
\end{eqnarray}
Combining this with the (average) product BF in eq.~\ref{eq.1}
yields a lower limit on the $X(3872)$ decay BF
\begin{eqnarray}
\BF(X(3872)\ra \pppm\jpsi) > 0.042\ @\ 90\%\ {\rm C.L.}
\label{eq.4}
\end{eqnarray}

\begin{figure}[h]
\centering
\includegraphics[width=80mm]{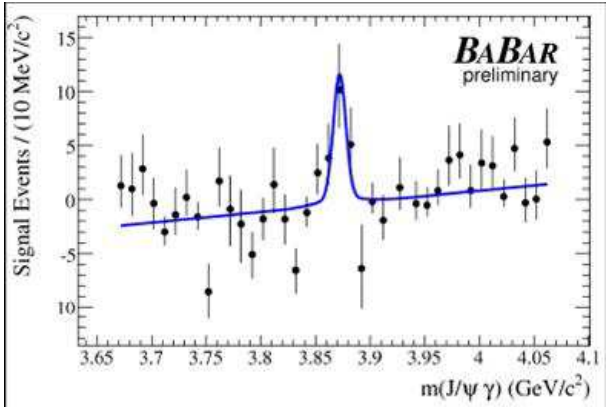}
\caption{Number of $B$ signal events in bins of $\gamma\,\jpsi$ invariant mass.} 
\label{fig.gammaJpsi}
\end{figure}

Recently, \babar\ confirmed Belle's observation~\cite{hepex.0505037} 
of the radiative decay
$X(3872)\ra \gamma\,\jpsi$. The existence of this decay mode determines the 
$C$-parity of the $X(3872)$ to be positive. 
Using 287M \BBbar\ events, \babar\ observes $19.4 \pm 5.7$ signal events;
see Fig.~\ref{fig.gammaJpsi}. 
The corresponding (preliminary) product BF is
\begin{eqnarray}
\BF(B\ra K\,X(3872),\ X(3872)\ra \gamma\,\jpsi) = \nonumber \\
(3.4 \pm 1.0 \pm 0.3) \times 10^{-6}\,,
\label{eq.5}
\end{eqnarray}
where the errors are statistical and systematic, respectively.
Combining this with Belle's measurement
yields
\begin{eqnarray}
\BF(B\ra K\,X(3872),\ X(3872)\ra \gamma\,\jpsi) = \nonumber \\
(2.2 \pm 0.7) \times 10^{-6}\,,
\label{eq.6}
\end{eqnarray}
where the error has been scaled by a factor $S=1.3$ according to the 
particle data group (PDG) prescription~\cite{pdg}. 
The resulting ratio of radiative over hadronic decays is
\begin{eqnarray}
\frac{\BF(X(3872)\ra \gamma\,\jpsi)}{\BF(X(3872)\ra \pppm\jpsi)} = 0.19 \pm 0.07\,.
\label{eq.7}
\end{eqnarray}

The positive $C$-parity of the $X(3872)$ fixes the QN of the 
\pppm\ system to be
${\rm I^G}(\JPC) = 1^+(1^{--})$, \ie, those of the $\rho$~meson.
Consequently, one expects the \pppm\ invariant mass spectrum to
peak toward the upper kinematic boundary. That is just what Belle
observes in their analysis~\cite{hepex.0505038}.
The \pppm\ invariant mass spectrum can also yield clues to the 
relative angular momentum, $L$, between the \pppm\ system and the \jpsi,
due to the phase-space suppression that scales with $[q^*_\jpsi]^{2L+1}$,
where $q^*_\jpsi$ is the \jpsi\ momentum in the $X(3872)$ rest frame.
The \pppm\ spectrum is fitted 
using a background function (estimated
from $X(3872)$ sidebands) plus a $\rho$ Breit-Wigner (B-W) function
modified by an S-wave (solid line) or P-wave (dashed line) phase-space
factor. The data clearly prefers S-wave; as a consequence positive parity
is preferred for the $X(3872)$~\footnote{Note that this conclusion is
significantly weakened if a more complicated decay dynamics, \eg, 
$\rho-\omega$ interference, is considered, as described in a CDF 
analysis~\cite{hepex.0512074}.}.
So at this point, the preferred $X(3872)$ spin-parity assignment is among
$\JPC = 0^{++},\ 1^{++},\ 2^{++}$.

A short digression on isospin here:
the positive $C$-parity of the $X(3872)$ and the 
resulting QNs of the \pppm\ system determine the $\pppm\jpsi$ final state
to be an isovector. Using 234M \BBbar\ pairs, \babar\ has searched for 
charged partners $X^\pm(3872)$ and determines an upper limit~\cite{prd71.031501}
\begin{eqnarray}
\BF(B^0\ra K^\mp\,X^\pm(3872),\ X^\pm(3872)\ra \pi^\pm\piz\jpsi) \nonumber\\
< 5.4 \times 10^{-6}\ @\ 90\%\ {\rm C.L.}\,,
\label{eq.8}
\end{eqnarray}
about a factor 2 lower than the BF observed for the neutral $X(3872)$. 
Thus, one concludes that the initial-state neutral $X(3872)$ is an isoscalar
and its decay into $\pppm\jpsi$ violates isospin symmetry.
It is this suppression that causes the very small $X(3872)$ width.
One might ask then: what prevents the isospin-allowed decay
$X(3872)\ra \pppm\piz\jpsi$ from happening all too frequently? 
Examining the QNs of a 
neutral 3-pion system resulting from a positive $C$-parity initial state,
one finds
${\rm I^G}(\JPC) = 0^-(1^{--})$, \ie, those of the $\omega$~meson.
However, at the nominal $\omega$~mass of $\sim 782\mevcc$, the $\omega\jpsi$
decay is kinematically forbidden. It is only due to the $\sim 8.5\mev$
natural width of the $\omega$ that this decay channel is accessible,
leaving it (kinematically) suppressed to a level comparable to the 
isospin-violating decay via the $\rho$:
\begin{eqnarray}
\frac{\BF(X(3872)\ra \pppm\piz\jpsi)}{\BF(X(3872)\ra \pppm\jpsi)}
= 1.0 \pm 0.4 \pm 0.3\,,
\label{eq.9}
\end{eqnarray}
as measured by Belle~\cite{hepex.0505037}.

Belle has extensively studied~\cite{hepex.0505038} 
the decay angular distributions in the decay
$B\ra K\,X(3872),\ X(3872)\ra \pppm\jpsi$, 
which is relatively straightforward thanks to the initial \Bmeson\ and the 
accompanying kaon being spin-less (pseudoscalar) particles.
Looking at a particular pair of $X(3872)$ decay angles suggested in ref.~\cite{prd70.094023},
they find their distributions to be entirely consistent with the expectation for
$\JPC = 1^{++}$, whereas there is poor consistency for the $\JPC = 0^{++}$
hypothesis. Unfortunately, the data is inclusive for $\JPC = 2^{++}$.

\begin{figure}[h]
\centering
\includegraphics[width=80mm]{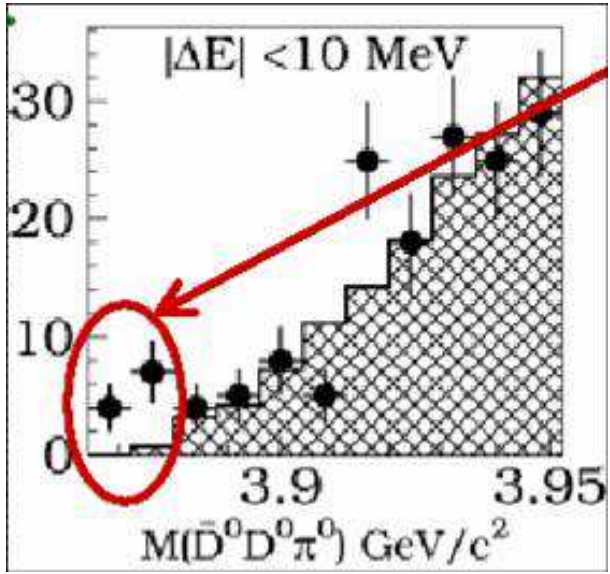}
\caption{$\bar D^0 D^0 \piz$ invariant mass distribution in $B$ signal region;
cross-hatched: $B$ sideband region.}
\label{fig.mD0barD0pi0}
\end{figure}

At a number of conferences earlier this year, Belle has reported (preliminary)
results on the observation of the $X(3872)\ra \bar D^0 D^0 \piz$ decay mode. Analyzing a
sample of 270M \BBbar\ events, they measure~\footnote{It should be noted that
this measurement is in poor agreement with a previously published~\cite{prl93.051803}
upper limit on this mode, 
$\BF(B\ra K\,X(3872),\ X(3872)\ra \bar D^0 D^0 \piz)
< 6 \times 10^{-5}$ @ 90\% C.L., 
based on an analysis of 96M \BBbar\ pairs. Resolution of 
this issue is awaiting a forthcoming publication on the new measurement.}
\begin{eqnarray}
\BF(B\ra K\,X(3872),\ X(3872)\ra \bar D^0 D^0 \piz) = \nonumber \\
(1.6 \pm 0.4 \pm 0.3) \times 10^{-4}\,,
\label{eq.10}
\end{eqnarray}
which is about an order of magnitude larger than the original discovery mode:
\begin{eqnarray}
\frac{\BF(X(3872)\ra \bar D^0 D^0 \piz )}{\BF(X(3872)\ra \pppm\jpsi)}
= 13.8 \pm 4.9\,.
\label{eq.11}
\end{eqnarray}
Fig.~\ref{fig.mD0barD0pi0} shows the
$\bar D^0 D^0 \piz$ invariant mass distribution, which exhibits an excess of
$12.5 \pm 3.9$ events in the vicinity of 3872\mevcc\ that is not present 
for events in the $B$ sideband region. 
The existence of this decay mode not only excludes 
$\JP = 0^{+}$ 
on the basis of spin-parity conservation, but it also argues strongly against a  
$\JP = 2^{+}$ 
assignment as that would require a D-wave angular momentum to be present, which is  
very unlikely for a system within $\sim 8\mevcc$ of threshold.
Unfortunately, statistics is not sufficient to determine whether this final state 
really originates from 
$X(3872)\ra \bar D^0 D^{*0}$ decay.

The $X(3872)$ has been searched for, but not found, in ISR production~\cite{prd71.052001}
\begin{eqnarray}
\Gamma^X_{ee} \times \BF(X(3872)\ra \pppm\jpsi) \nonumber \\
< 6.2\ev\ @\ 90\%\ {\rm C.L.}\,,
\label{eq.12}
\end{eqnarray}
confirming that it is not a $\JPC = 1^{--}$ state.
Similarly, no sign of it was found in quasi-real two-photon collisions~\cite{prl94.032004}
\begin{eqnarray}
\Gamma^X_{\gg} \times \BF(X(3872)\ra \pppm\jpsi) \nonumber \\
< \frac{12.9\ev}{2\rm J + 1}\ @\ 90\%\ {\rm C.L.}\,,
\label{eq.13}
\end{eqnarray}
confirming that it is not a spin-0 or spin-2 state.

In summary, the $X(3872)$ is a very narrow $\JPC = 1^{++}$ state with a mass
right at $\bar D^0 D^{*0}$ threshold (indistinguishable within errors), 
decaying dominantly into
$\bar D^0 D^0 \piz$. Its properties are perfectly consistent with the interpretation
as a $\bar D^0$-$D^{*0}$ molecule, whereas it is very difficult to accommodate as a 
regular charmonium state.

\subsection{$\bf X(3940)$}

Using 375\invfb\ of data, Belle has investigated double-charmonium production
in \ccbar\ continuum events~\cite{hepex.0507019}. 
They examine the invariant mass of the system recoiling against a reconstructed
\jpsi\ and observe a structure near 3940\mevcc, containing $266 \pm 63$ events. 
This $X(3940)$ is produced together with three
known charmonium states: $\eta_c$, $\chi_{c0}$, and $\eta_c(2S)$ (all of which 
have positive $C$-parity). A very natural interpretation of this structure is
to identify it as the $\eta_c(3S)$, which would make it a 
$\JPC = 0^{-+}$ (pseudo-scalar) state.
A charmonium state of this mass is expected to decay dominantly into
open charm. Belle has searched for its $D\bar D$ and $D^*\bar D$ decay
by constraining the recoil mass against a reconstructed $\jpsi \bar D$
system to be either around the $D$ or $D^*$ mass and examining the 
mass spectrum of the system recoiling against the \jpsi. In the $D^* \bar D$
case, they observe a $X(3940)$ signal of $24.5 \pm 6.9$ events with a mass
and width
\begin{eqnarray}
m_X      & = & (3943 \pm 6 \pm 6)\mevcc, \nonumber \\
\Gamma_X & = & (15.4 \pm 10.1_{stat})\mev \nonumber \\
         & < & 52\mev\ @\ 90\%\ {\rm C.L.}
\label{eq.14}
\end{eqnarray}
The corresponding BF is
\begin{eqnarray}
\BF(X(3940)\ra D^* \bar D) & = & 0.96^{+0.04}_{-0.32}\pm 0.22 \nonumber \\
                     & > & 0.45\ @\ 90\%\ {\rm C.L.}
\label{eq.15}
\end{eqnarray}
No signal is observed in the $D \bar D$ case, resulting in an upper limit
\begin{eqnarray}
\BF(X(3940)\ra D \bar D) & < & 0.41\ @\ 90\%\ {\rm C.L.}
\label{eq.16}
\end{eqnarray}
The absence of a $D\bar D$ decay for the $X(3940)$ is consistent
with its assignment as the $\eta_c(3S)$, since a pseudo-scalar cannot
decay into two pseudo-scalars.

Belle has also searched for the decay $X(3940)\ra \omega\jpsi$,
determining an upper limit
\begin{eqnarray}
\BF(X(3940)\ra \omega\jpsi) & < & 0.29\ @\ 90\%\ {\rm C.L.}
\label{eq.17}
\end{eqnarray}
The motivation for this will become clear in the next section.

\subsection{$\bf Y(3940)$}

Analyzing 275M \BBbar\ decays, Belle has observed a large $\omega\jpsi$ 
threshold enhancement~\cite{prl94.182002} in the channel
$B^\pm \ra K^\pm\ \omega\jpsi$.
If interpreted as a S-wave resonance, a premise that should be critically
reviewed, a B-W fit yields a signal of $58 \pm 11$ events with a mass and width 
\begin{eqnarray}
m_Y      & = & (3943 \pm 11 \pm 13)\mevcc, \nonumber \\
\Gamma_Y & = & (87 \pm 22 \pm 26)\mev \nonumber \\
         & > & 20\mev\ @\ 90\%\ {\rm C.L.}
\label{eq.18}
\end{eqnarray}
(The mechanics of this fit should also be reviewed; I find it surprising 
that a MINOS error evaluation would yield symmetric errors in this (threshold)
situation.)
The observed number of signal events corresponds to a BF
\begin{eqnarray}
\BF(B^\pm \ra K^\pm\,Y(3940),\ Y(3940)\ra \omega\jpsi) \nonumber \\
= (7.3 \pm 1.3 \pm 3.1) \times 10^{-5}\,.
\label{eq.19}
\end{eqnarray}

If this threshold enhancement is really a resonance state, its nature
is totally unclear. 
A charmonium state of this mass is expected to decay dominantly into
open charm; $\omega\jpsi$ is not a decay mode that readily comes to mind.
That decay mode 
fixes the $Y(3940)$ $C$-parity to be positive---just like for the $X(3940)$ 
discussed in the previous section.
Given the weak upper limit for the $X(3940)\ra \omega\jpsi$ decay 
and the large uncertainties in the $X(3940)$ and $Y(3940)$ widths,
it is not clear (to me) whether these really are different states.

\subsection{$\bf Z(3930)$}

In an effort to elucidate the nature of the $X/Y$ states discussed in the
previous sections, Belle has searched for $D^0\bar D^0$ and $D^+D^-$ production
in two-photon interactions~\cite{prl96.082003}, using 395\invfb\ of data.
As mentioned before, the production mechanism results in positive $C$-parity
final states with spin 0 or spin 2. In case of spin 2, there are two possible
helicity states, 0 and 2, with helicity-2 : helicity-0 = 6 : 1 due to Clebsch-Gordon
coefficients.
Belle observes a signal of $64 \pm 18$ events with mass and width
\begin{eqnarray}
m_Z      & = & (3929 \pm 5 \pm 2)\mevcc, \nonumber \\
\Gamma_Z & = & (29 \pm 10 \pm 2)\mev\,.
\label{eq.20}
\end{eqnarray}
To determine the spin of this state, they investigate, in the $Z(3940)$ signal region,
the angular distribution of one of the $D$s with respect to (w.r.t.) the beam axis in the
$\gg$~CM frame, and find it to be entirely consistent with spin 2, helicity 2,
whereas spin 0 is strongly disfavored. Using this spin and helicity assignment,
they measure
\begin{eqnarray}
\Gamma^Z_{\gg} \times \BF(Z(3930)\ra D\bar D) \nonumber \\
= (0.18 \pm 0.05 \pm 0.03)\kev\,.
\label{eq.21}
\end{eqnarray}
Such a two-photon partial width (assuming a dominant decay to open charm)
is in reasonable agreement with the expectation~\cite{prd69.094019}
for a conventional charmonium state at this mass.
Therefore, the $Z(3940)$ is plausibly identified as a $2\, ^3\!P_2$ \ccbar\ state, 
the $\chi_{c2}(2P)$.

\subsection {$\bf Y(4260)$}

Searching for the $X(3872)$ in ISR events in a data sample of 233\invfb, 
\babar\ observed a broad structure in the \pppm\jpsi\ mass spectrum 
around 4.26\gevcc, the $Y(4260)$~\cite{prl95.142001}. 
The ISR production mechanism fixes the QNs of the final state to be $\JPC = 1^{--}$. 
Assuming that the observed structure is a single resonance, a B-W fit yields
$125 \pm 23$ signal events corresponding to an electronic width times BF
\begin{eqnarray}
\Gamma^Y_{ee} \times \BF(Y(4260)\ra \pppm\jpsi) \nonumber \\
= (5.5 \pm 1\, ^{+0.8}_{-0.7})\ev\,,
\label{eq.22}
\end{eqnarray}
a peak cross section~\cite{prd73.012005}
\begin{eqnarray}
\sigma(\epem \ra Y(4260)) = (51 \pm 12)\pb\,,
\label{eq.23}
\end{eqnarray}
and a mass and width
\begin{eqnarray}
m_Y      & = & (4259 \pm 8\, ^{+2}_{-6})\mevcc, \nonumber \\
\Gamma_Y & = & (88 \pm 23\, ^{+6}_{-4})\mev\,.
\label{eq.24}
\end{eqnarray}

Up to very recently, the only other sighting of the $Y(4260)$ 
came from a \babar\ measurement in \Bdcy~\cite{prd73.011101}, 
where a weak ($\sim 3\,\sigma$) signal was observed in 
\begin{eqnarray}
\BF(B^\pm \ra K^\pm\,Y(4260),\ Y(4260)\ra \pppm\jpsi) \nonumber \\
= (2.0 \pm 0.7 \pm 0.2) \times 10^{-5}\,.
\label{eq.25}
\end{eqnarray}
It would be important to repeat this measurement, which used ``only''
232M \BBbar\ events, with a larger data sample.

Fortunately, definitive confirmation of the $Y(4260)$ just arrived in the
form of a CLEO-c scan~\cite{prl96.162003}
of the CM energy region $\sqrt{s} = 3.77-4.26\gev$.
At the highest scan point, $\sqrt{s} = 4.26\gev$, where they collected
13.2\invpb\ of data, they observe an enhanced cross section for
$\epem\ra \pi\pi\jpsi$, for charged as well as for neutral pions
\begin{eqnarray}
\sigma(\epem \ra \pppm\jpsi)    = (58\, ^{+12}_{-10} \pm 4)\nb\,, \nonumber \\
\sigma(\epem \ra \piz\piz\jpsi) = (23\, ^{+12}_{-8} \pm 1)\nb\,.
\label{eq.26}
\end{eqnarray}
Even though the two measurements are based on only 37 and 8 events, respectively,
they are highly significant, thanks to very low backgrounds. Note that the charged-pion
measurement compares very favorably with the \babar\ result in eq.~\ref{eq.23}.
Since the $C$-parity of the $Y(4260)$ is negative, the QNs of the $\pi\pi$ system
are ${\rm I^G}(\JPC) = 0^+(0^{++},\ 2^{++})$. Thus, $\pppm$ and $\piz\piz$ are
expected to occur with a (isospin) ratio of 2:1, in agreement with the CLEO measurement
in eq.~\ref{eq.26}.  
CLEO observes no sign of an enhanced $\pi\pi\jpsi$ decay of the $\psi(4040)$, 
which is considered to be the $\psi(3S)$ state; this makes the assignment of the
$Y(4260)$ as the $\psi(4S)$ state less likely.
At this conference, CLEO also showed (preliminary) results from a search for
ISR-produced $Y(4260)\ra \pppm\jpsi$ using 13.3\invfb\ of CLEO-III data 
collected on or near the \FourS\ resonance. 
They observe a clear $Y(4260)$ signal of 12 events with very little background.

\begin{figure}[h]
\centering
\includegraphics[width=80mm]{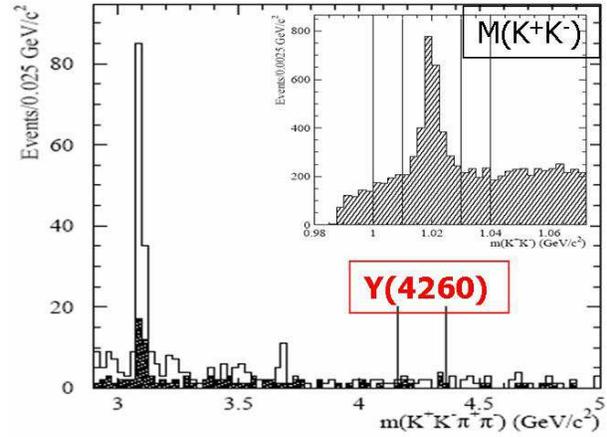}
\caption{$\pppm K^+K^-$ invariant mass spectrum; $\phi$ sidebands are shaded.
The inset shows the $K^+K^-$ invariant mass, with $\phi$ signal and sideband
regions indicated by the vertical lines.}
\label{fig.pipiphi}
\end{figure}

\babar\ has searched for the $Y(4260)$ in other ISR-produced final states,
\ie, $\pppm\phi$, $p\bar p$, and $D\bar D$, but observes no signal in any
of those.
Fig.~\ref{fig.pipiphi} shows the $\pppm K^+K^-$ invariant mass spectrum
from 232\invfb\ of data; no significant structure other than the \jpsi\
(and maybe a hint of the $\psi(2S)$) is observed, resulting in the (preliminary)
upper limit
\begin{eqnarray}
\Gamma^Y_{ee} \times \BF(Y(4260)\ra \pppm\phi) \nonumber \\
< 0.4\ev\ @\ 90\%\ {\rm C.L.}
\label{eq.27}
\end{eqnarray}

The same dataset has been used to investigate the $p\bar p$ final state~\cite{prd73.012005}:
\begin{eqnarray}
\frac{\BF(Y(4260)\ra p\bar p)}{\BF(Y(4260)\ra \pppm\jpsi)}
< 0.13\ @\ 90\%\ {\rm C.L.}
\label{eq.28}
\end{eqnarray}

\begin{figure}[h]
\centering
\includegraphics[width=80mm]{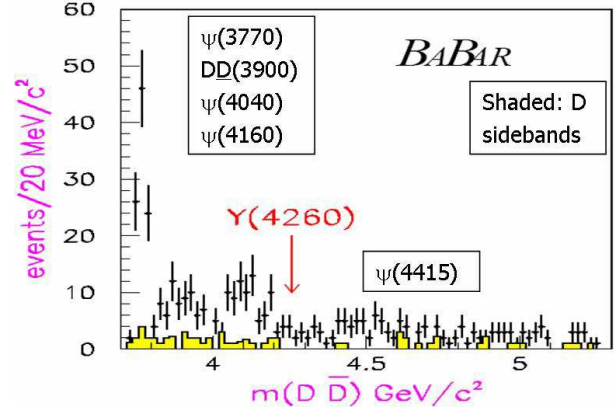}
\caption{$D \bar D$ invariant mass spectrum; $D$ sidebands are shaded yellow.}
\label{fig.DDbar}
\end{figure}

Finally, using 289\invfb\ of data, \babar\ has measured ISR production of $D\bar D$.
Fig.~\ref{fig.DDbar} shows the $D\bar D$ invariant mass spectrum, which shows a
significant $\psi(3770)$ signal; it also exhibits several structures 
previously observed in \epem\ $R$-scans which are candidates for 
$\JPC = 1^{--}$ \ccbar\ states, for example $\psi(4040)$, $\psi(4160)$, 
and $\psi(4415)$. The one place, however, where no structure is apparent
in the \babar\ spectrum is at a $D\bar D$ mass of 4260\mevcc.
The corresponding (preliminary) upper limit is, unfortunately, not very stringent due to the
very low efficiency for reconstructing two $D$ mesons 
\begin{eqnarray}
\frac{\BF(Y(4260)\ra D\bar D)}{\BF(Y(4260)\ra \pppm\jpsi)}
< 7.6\ @\ 95\%\ {\rm C.L.}
\label{eq.29}
\end{eqnarray}

In summary, the existence of the $Y(4260)$ is now well established, as are
its $\JPC = 1^{--}$ QNs. However, its nature remains a complete mystery:
there is no room for it in the spectrum of $1^{--}$ charmonium states,
and the only final state it has been observed in, $\pi\pi\jpsi$, is a 
rather unexpected one for a conventional \ccbar\ state of this mass.  
Exciting unconventional explanations for the $Y(4260)$ have been suggested, \eg, 
as a $\ccbar g$ hybrid state, that will be examined as the data samples increase. 

\section{Charmed-strange Mesons}

The 
$D^*_{sJ}(2317)$ 
and 
$D_{sJ}(2460)$ 
were first observed by \babar~\cite{prl90.242001}
and CLEO~\cite{prd68.032002} in \ccbar\ continuum events, and by Belle~\cite{prl91.262002}
in \Bdcy, using data samples of 91\invfb, 13.5\invfb, and 124M \BBbar\ events, respectively.
The masses of both states are unexpectedly low: below $DK$ and $D^*K$ threshold,
respectively. As a consequence, only isospin-violating or electromagnetic decays are
kinematically allowed, resulting in very narrow widths for both states.
Apart from their low masses, the
$D^*_{sJ}(2317)$ 
and 
$D_{sJ}(2460)$ 
decay patterns and angular distributions are consistent with their interpretation
as conventional P-wave $c\bar s$ mesons with $\JP = 0^+$ and $\JP = 1^+$, respectively. 

\begin{figure}[h]
\centering
\includegraphics[width=80mm]{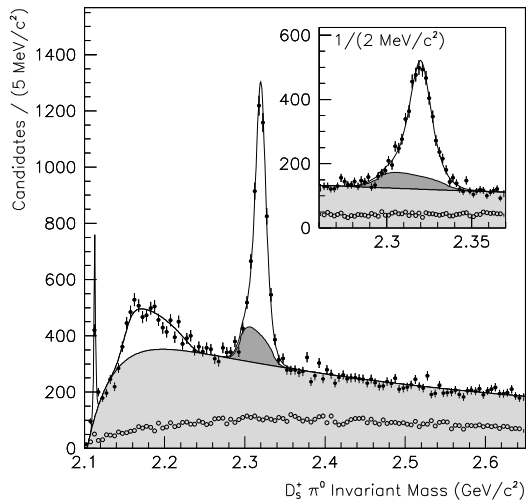}
\caption{
The invariant mass
distribution for (solid points) $\Ds\piz$ candidates and (open points)
the equivalent using the $\Ds$ sidebands.
The curve represents the likelihood fit described in
the text. Included in this fit is (light shade) a contribution
from combinatorial background and (dark shade) the reflection from
$\DsFE\to\DsTO\piz$ decay. The insert highlights
the details near the $\DsTT$ mass.
}
\label{fig.DsJ2317.Dspi0} 
\end{figure}

\babar\ has updated their analysis to use a data sample of 232\invfb\ 
and performed a comprehensive study~\cite{DsJ2317and2460}
of \DsJ\ decays to \Ds\ plus one or two charged pions, neutral pions, or photons;
a total of 8 and 9 modes for 
$D^*_{sJ}(2317)$ 
and 
$D_{sJ}(2460)$, 
respectively, covering all such channels---allowed by spin-parity
conservation or not---that are kinematically allowed. 
The 
$D^*_{sJ}(2317)^+$ 
is seen in only one mode: the (allowed) discovery mode
$D^*_{sJ}(2317)^+ \ra \Ds \piz$;
searches in all other modes yield only upper limits.
Fig.~\ref{fig.DsJ2317.Dspi0} shows the $\Ds\piz$ invariant mass spectrum.
It shows a very large 
$D^*_{sJ}(2317)^+$ signal (${\cal O}(3000)$ events) 
plus a 
$D_s^*(2112)^+)$
signal as well as
reflections from the 
$D_s^*(2112)^+)$
and the
$D_{sJ}(2460)^+$.
The latter reflection piles up right underneath the  
$D^*_{sJ}(2317)^+$ 
signal, making the extraction of yield and resonance parameters 
quite difficult. After a detailed study of the shapes of the
various contributions to the spectrum, \babar\ measures the 
$D^*_{sJ}(2317)^+$ 
%raw yield to be
%$3180 \pm 80_{stat}$ events, and its
mass and width
\begin{eqnarray}
m      & = & (2319.6 \pm 0.2 \pm 1.4)\mevcc, \nonumber \\
\Gamma & < & 3.8\mev\ @\ 95\%\ {\rm C.L.}
\label{eq.30}
\end{eqnarray}

They have searched for neutral or doubly-charged partners of the
$D^*_{sJ}(2317)^+$ in the
$\Ds\pi^-$ and  $\Ds\pi^+$ channel,  
but find no indication of such states,
thus concluding that the 
$D^*_{sJ}(2317)$ 
is an isoscalar.

\begin{figure}[h]
\centering
\includegraphics[width=80mm]{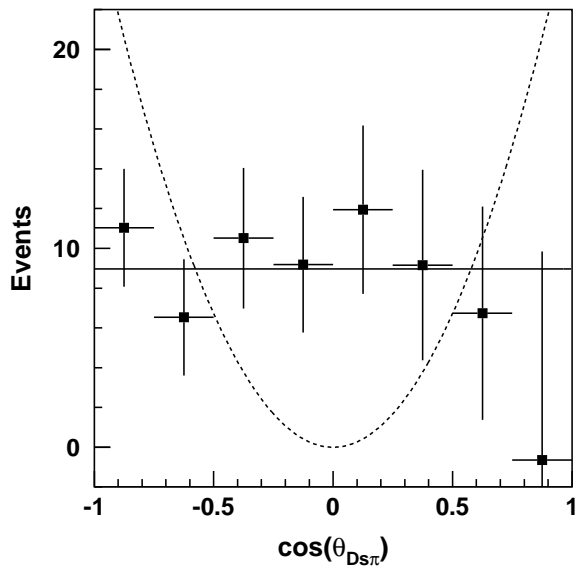}
\caption{Efficiency corrected angular distribution for $D^*_{sJ}(2317)^+\ra \Ds\piz$. 
The solid (dashed) line shows the expectation for spin 0 (spin 1).
}
\label{fig.DsJ2317.costht} 
\end{figure}

Using 274M \BBbar\ events, Belle has studied~\cite{belle-conf-0461}
the decay angular distributions in 
$B\ra \bar D D^*_{sJ}(2317)^+,\, D^*_{sJ}(2317)^+\ra \Ds\piz$, 
by looking at the angle of the \Ds\ w.r.t. the 
$D^*_{sJ}(2317)^+$ 
flight direction in the 
$D^*_{sJ}(2317)^+$ 
CM frame, 
$\theta_{D_s\pi}$; see Fig.~\ref{fig.DsJ2317.costht}.
They find this angle to have a flat distribution,
consistent with spin 0 and inconsistent with the 
$dN/d\cos\theta_{D_s\pi} \propto \cos^2\theta_{D_s\pi}$ 
expectation for spin 1. 
So from the observed decay mode and angular distribution,
one concludes that the $D^*_{sJ}(2317)$ is a $\JP = 0^+$ particle.

\begin{figure}[h]
\centering
\includegraphics[width=80mm]{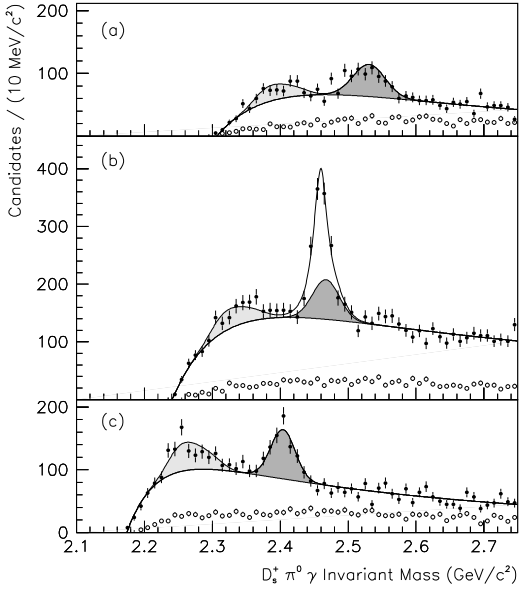}
\caption{
The invariant mass distribution
of $\Ds\piz\gamma$ candidates in the (a) upper, (b) signal,
and (c) lower $\Ds\gamma$ mass selection windows for (solid points)
the $\Ds$ signal and (open points) $\Ds$ sideband samples. 
The dark gray (light gray)
region corresponds to the predicted contribution from the
$\DsTT$ ($\DsTO$) reflection.
}
\label{fig.DsJ2460.Dspi0gamma} 
\end{figure}
\begin{figure}[h]
\centering
\includegraphics[width=80mm]{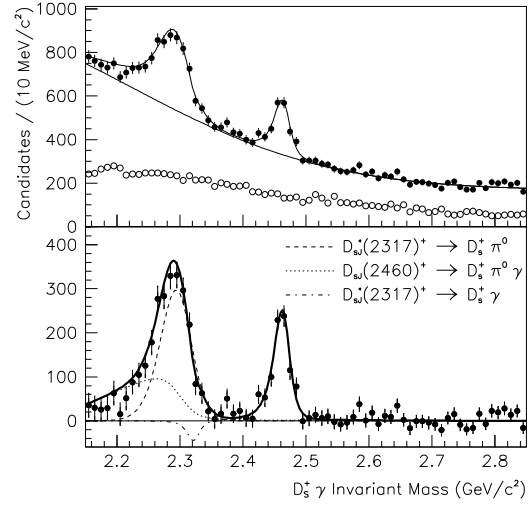}
\caption{
$\Ds\gamma$ invariant mass distribution. The solid points in the top
plot are the mass distribution. The open points are the $\Ds$ sidebands,
scaled appropriately. The bottom plot shows the same data 
after subtracting the background curve from the fit. Various contributions
to the likelihood fit are also shown.
}
\label{fig.DsJ2460.Dsgamma} 
\end{figure}
\babar\ observes the
$D_{sJ}(2460)$ 
in three channels:
$D_{sJ}(2460)^+ \ra D^*_s(2112)^+ \piz$ with $D^*_s(2112)^+\ra \Ds\gamma$,
$D_{sJ}(2460)^+ \ra \Ds\gamma$,
and
$D_{sJ}(2460)^+ \ra \Ds\pppm$.
Fig.~\ref{fig.DsJ2460.Dspi0gamma} shows the 
$\Ds\piz\gamma$ invariant mass spectrum.
It shows a large
$D_{sJ}(2460)^+$ signal (${\cal O}(600)$ events)  
as well as reflections from the 
$D_s^*(2112)^+$
and the
$D^*_{sJ}(2317)^+$.
The latter reflection piles up right underneath the  
$D_{sJ}(2460)^+$
signal, making the extraction of yield, mass, and width
quite difficult. 
%After a detailed study of the shapes of the
%various contributions to the spectrum, \babar\ measures the 
%$D_{sJ}(2460)^+$ raw yield to be
%$560 \pm 40_{stat}$ events, and its
%mass and width
%\begin{eqnarray}
%m      & = & (2458.6 \pm 1.0 \pm 2.5)\mevcc, \nonumber \\
%\Gamma & < & 6.3\mev\ @\ 95\%\ {\rm C.L.}
%\label{eq.30}
%\end{eqnarray}
Moreover, the 
resonance parameters 
can be determined with much better precision in the (all-charged)
\Ds\pppm\ final state; see below.

Fig.~\ref{fig.DsJ2460.Dsgamma} shows the 
$\Ds\gamma$ invariant mass spectrum.
It shows a large
$D_{sJ}(2460)^+$ signal (${\cal O}(900)$ events)  
as well as reflections from the 
$D^*_{sJ}(2317)^+$ and the $D_{sJ}(2460)^+$ itself.
This time, though, the reflections produce peaks well below the 
$D_{sJ}(2460)^+$.
\babar\ measures the ratio of BFs
\begin{eqnarray}
\frac{\BF(D_{sJ}(2460)^+\ra \Ds\gamma)}{\BF(D_{sJ}(2460)^+\ra \Ds\piz\gamma)} \nonumber \\
= 0.337 \pm 0.036 \pm 0.038\,.
\label{eq.31}
\end{eqnarray}
Note that the existence of this decay mode rules out spin 0 for the $D_{sJ}(2460)$. 

Belle has studied~\cite{belle-conf-0461}
the $B\ra \bar D D_{sJ}(2460)^+$ 
decay angular distributions for 
$D_{sJ}(2460)^+\ra \Ds\gamma$ as well as  
$D_{sJ}(2460)^+\ra D^*_s(2112)^+\piz$
by looking at the angle of the \Ds\ ($D^*_s(2112)^+$) w.r.t. the 
$D_{sJ}(2460)^+$ 
flight direction in the 
$D_{sJ}(2460)^+$ 
CM frame, 
$\theta_{D_s\gamma}$ ($\theta_{D^*_s\pi}$).
\begin{figure}[h]
\centering
\includegraphics[width=80mm]{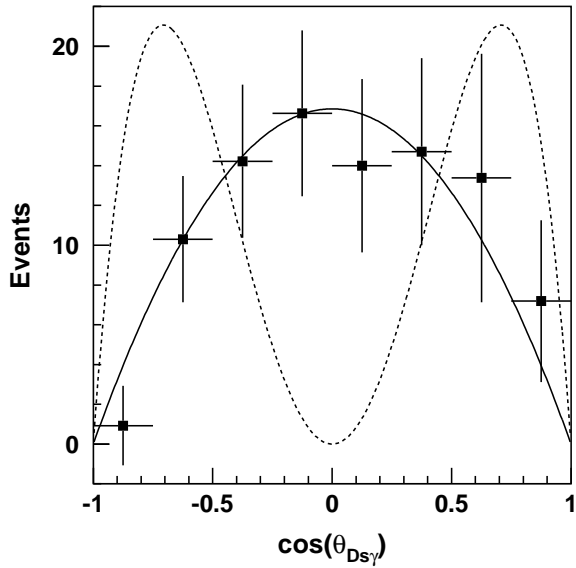}
\caption{Efficiency corrected angular distribution, 
$\cos \theta_{D_s\gamma}$,
for 
$D_{sJ}(2460)^+\ra \Ds\gamma$.
The solid (dashed) line shows the expectation for 
spin 1 (spin 2).
}
\label{fig.DsJ2460a.costht} 
\end{figure}
\begin{figure}[h]
\centering
\includegraphics[width=80mm]{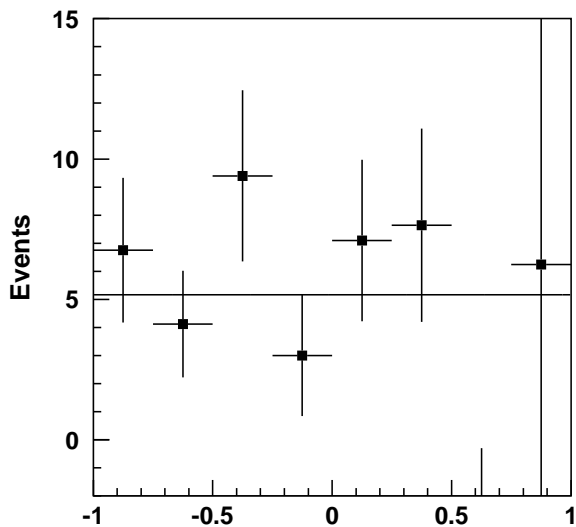}
\caption{Efficiency corrected angular distribution, 
$\cos\theta_{D^*_s\pi}$,
for 
$D_{sJ}(2460)^+\ra D^*_s(2112)^+\piz$.
The solid line shows the expectation for 
spin 1 and positive parity, for a pure S-wave.
}
\label{fig.DsJ2460b.costht} 
\end{figure}
As shown in Fig.~\ref{fig.DsJ2460a.costht} for the
$\Ds\gamma$ final state, the distribution is perfectly consistent
with the spin-1 expectation, 
$dN/d\cos\theta_{D_s\gamma} \propto 1 - \cos^2\theta_{D_s\gamma}$, 
and totally inconsistent with the spin-2 expectation,
$dN/d\cos\theta_{D_s\gamma} \propto \sin^2\theta_{D_s\gamma} \cos^2\theta_{D_s\gamma}$.
Note that the angular distributions in this decay mode cannot distinguish
between positive and negative parity; this is a consequence of the (massless) 
photon missing the helicity-0 state.
Having established that the $D_{sJ}(2460)$ is a spin-1 particle,
one can use the $D^*_s(2112)^+\piz$ final state to establish its parity
(with the $D^*_s(2112)$ being a (massive) vector particle with all three
helicity components):
the $\theta_{D^*_s\pi}$ distribution (see Fig.~\ref{fig.DsJ2460b.costht}) 
is consistent with being flat, which is the expectation for a $\JP = 1^+$
state with pure S-wave between the $D^*_s(2112)^+$ and the \piz\ (though
the appropriate combination of S- and D-wave could also produce a flat 
distribution). 
More importantly, the data is inconsistent with the expectation for $\JP = 1^-$,
which is
$dN/d\cos\theta_{D^*_s\piz} \propto 1 - \cos^2\theta_{D^*_s\piz}$. 
So from the observed decay modes and angular distributions,
one concludes that the $D_{sJ}(2460)$ is a spin-1 particle with positive parity.

\begin{figure}[h]
\centering
\includegraphics[width=80mm]{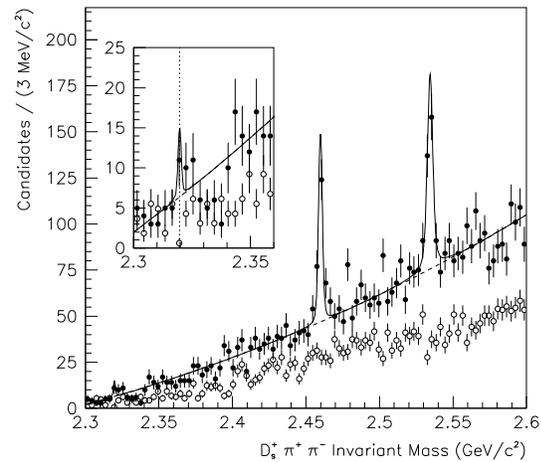}
\caption{
The invariant mass distribution
of (solid points) $\Ds\pppm$ candidates
and (open points) the equivalent using the $\Ds$ sidebands. 
The insert focuses on the low mass region.
The dotted line in the insert indicates the $\DsTT$ mass.
}
\label{fig.DsJ2460.Dspipi} 
\end{figure}
Fig.~\ref{fig.DsJ2460.Dspipi} shows the 
$\Ds\pppm$ invariant mass spectrum.
It shows a sizable
$D_{sJ}(2460)^+$ signal (${\cal O}(100)$ events)  
as well as a similar-size 
$D_{s1}(2536)^+$ signal; there is no indication of the 
$D^*_{sJ}(2317)^+$.
Since this is an all-charged final state, \babar\ obtains
rather precise determinations of the mass and width of both states:
\begin{eqnarray}
m      & = & (2460.2 \pm 0.2 \pm 0.8)\mevcc, \nonumber \\
\Gamma & < & 3.5\mev\ @\ 95\%\ {\rm C.L.}\,,
\label{eq.32}
\end{eqnarray}
and
\begin{eqnarray}
m      & = & (2534.6 \pm 0.3 \pm 0.7)\mevcc, \nonumber \\
\Gamma & < & 2.5\mev\ @\ 95\%\ {\rm C.L.}
\label{eq.33}
\end{eqnarray}
They also measure the ratio of BFs
\begin{eqnarray}
\frac{\BF(D_{sJ}(2460)^+\ra \Ds\pppm)}{\BF(D_{sJ}(2460)^+\ra \Ds\piz\gamma)} \nonumber \\
= 0.077 \pm 0.013 \pm 0.008\,.
\label{eq.34}
\end{eqnarray}

No other $D_{sJ}(2460)$ decay channels are observed.
Tab.~\ref{tab.DsJBFs} gives a summary of the \babar\
$D^*_{sJ}(2317)$ 
and 
$D_{sJ}(2460)$ 
branching-ratio results. 

\begin{table}
\begin{center}
\caption{
A summary of branching-ratio results.
The first quoted uncertainty for the central value
is statistical and the second is systematic. The limits
correspond to 95\% CL. 
A lower limit is quoted for the $\DsFE\to\DsTO\piz$ results.
}
%\begin{ruledtabular}A
\renewcommand{\baselinestretch}{1.3}
\begin{tabular}{llr@{$\:\pm\:$}r@{$\:\pm\:$}rr}
%\begin{tabular}{llrrrr}
\hline\hline 
\multicolumn{2}{l}{Decay Mode} 
     & \multicolumn{3}{c}{Central Value} & Limit  \\
\hline
\multicolumn{6}{l}{$\mathcal B(\DsTT\to X)/\mathcal B(\DsTT\to\Ds\piz)$} \\
\hspace{5pt}
& $\Ds\gamma$        & $-0.02$&$ 0.02$&$ 0.08$&$ <0.14$\\
& $\Ds\pi^0\pi^0$    & $ 0.08$&$ 0.06$&$ 0.04$&$ <0.25$\\
& $\Ds\gamma\gamma$  & $ 0.06$&$ 0.04$&$ 0.02$&$ <0.18$\\
& $\DsTO\gamma$      & $ 0.00$&$ 0.03$&$ 0.07$&$ <0.16$\\
& $\Ds\pi^+\pi^-$    & $ 0.0023$&$ 0.0013$&$ 0.0002$&$ <0.0050$\\
\hline
\multicolumn{6}{l}{$\mathcal B(\DsFE\to X)/\mathcal B(\DsFE\to\Ds\piz\gamma)$ [a]} \\
& $\Ds\piz$          & $-0.023$&$ 0.032$&$ 0.005$&$ <0.042$ \\
& $\Ds\gamma$        & $ 0.337$&$ 0.036$&$ 0.038$&    ---   \\
& $\DsTO\piz$        & $ 0.97$&$ 0.09$&$ 0.05$&$ >0.75$\\
& $\DsTT\gamma$      & $ 0.03$&$ 0.09$&$ 0.05$&$ <0.25$\\
& $\Ds\pi^0\pi^0$    & $ 0.13$&$ 0.13$&$ 0.06$&$ <0.68$\\
& $\Ds\gamma\gamma$  & $ 0.08$&$ 0.10$&$ 0.04$&$ <0.33$\\
& $\DsTO\gamma$      & $-0.02$&$ 0.08$&$ 0.10$&$ <0.24$\\
& $\Ds\pi^+\pi^-$    & $ 0.077$ & $0.013$  & $0.008$ & ---\\
\hline 
\multicolumn{6}{l}{\footnotesize [a] Denominator includes both 
$D_s^*(2112)^+\piz$ and $D_{sJ}^*(2317)^+\gamma$\hfill}\\
\multicolumn{6}{l}{\footnotesize channels.\hfill}\\
\hline\hline 
\end{tabular}
%\end{ruledtabular}
\label{tab.DsJBFs}
\end{center}
\end{table}

\begin{figure}[h]
\centering
\includegraphics[width=80mm]{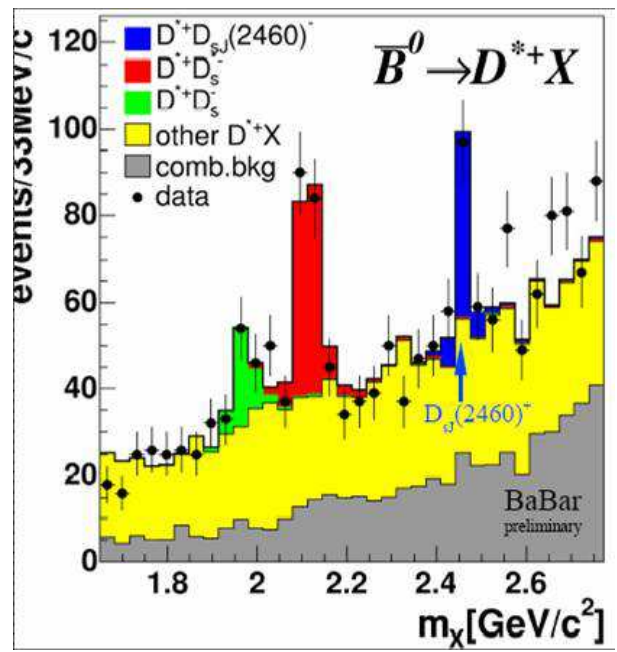}
\caption{Recoil mass, $m_x$, in the decay $\bar B ^0\ra D^{*+}\,X$ with
signal and background overlayed.
}
\label{fig.DsJ2460.mX} 
\end{figure}

\babar\ has determined~\cite{DsJ2460Abs}
for the first time absolute BFs for the $D_{sJ}(2460)$
using a sample of 230M \BBbar\ events, where they fully reconstruct
one \Bmeson; from the decay of the other $B$ they reconstruct a 
charged or neutral $D$ or $D^*$ and examine the mass of the $X$-system
recoiling against it. They observe $D_{sJ}(2460)$ signals in the recoil mass,
$m_X$, an example of which is shown in Fig.~\ref{fig.DsJ2460.mX} for a particular
\Bdcy.
Combining these results with previously measured~\cite{prl93.181801}
exclusive product BFs
$B\ra \bar D^{(*)} D_{sJ}(2460)^+,\ D_{sJ}(2460)^+\ra D^*_s(2112)^+\piz\,/\,D^+_s\gamma$
one obtains absolute BFs 
\begin{eqnarray}
\BF(D_{sJ}(2460)^+\ra D^*_s(2112)^+\piz & = & 0.56 \pm 0.13 \pm 0.09\,, \nonumber \\
\BF(D_{sJ}(2460)^+\ra D^+_s\gamma       & = & 0.16 \pm 0.04 \pm 0.03\,, \nonumber \\
\BF(D_{sJ}(2460)^+\ra D^+_s\pppm        & = & 0.04 \pm 0.01\,.
\label{eq.35}
\end{eqnarray}
Thus, the sum of all known $D_{sJ}(2460)$ BFs comes to $0.76 \pm 0.17$.
The error on this is unfortunately large: within $1.5\,\sigma$ we might 
conclude that all 
$D_{sJ}(2460)$ 
decays have been measured or that a large fraction
is still missing (although it is totally unclear what sizable channel could
be missing from the list that \babar\ investigated). More data will tell, 
and more data will eventually enable us to perform the same type of measurement
for the 
$D^*_{sJ}(2317)$.

In summary, \babar\ has obtained precise measurements of the $D^*_{sJ}(2317)$ and
$D_{sJ}(2460)$ masses, widths, and decay modes, including for the first time
absolute BFs, and Belle has determined the spin and parity of both states
(although it is most unfortunate that the analysis in ref.~\cite{belle-conf-0461}
has never been published). 
Most experimental data points to an interpretation as conventional P-wave $c\bar s$
mesons with $\JP = 0^+$ and $\JP = 1^+$, respectively. 
Two wrinkles remain in this picture: a lack of understanding why the masses are
lower than expected, and the apparent absence (at the current level of sensitivity) 
of certain radiative decays, \eg, 
$D^*_{sJ}(2317)^+\,/\,D_{sJ}(2460)^+ \ra D^*_s(2112)^+\gamma$.  
More theoretical work is needed to iron this out.

\section{Charmed Baryons}

There has been a lot of progress in charmed-baryon spectroscopy in recent years.
All nine $\JP = 1/2^+$ ground states ($L = 0$)
of singly-charmed baryons in the SU(4) 20'-plet, 
and all but one of six of the corresponding $\JP = 3/2^+$ 20-plet 
ground states have been observed. Furthermore, there is a growing number of 
excited ($L = 1$) states. 
There are not yet any confirmed sightings of 
doubly-charmed baryons.

\subsection{$\bf \Lambda_c(2940)$}

\begin{figure}[h]
\centering
\includegraphics[width=80mm]{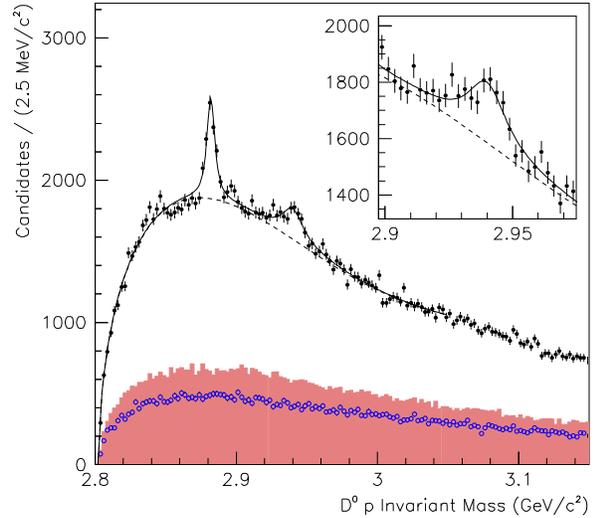}
\caption{
The solid points are the $D^0 p$ invariant mass distribution
of the final sample. Also shown are (shaded) the contribution from
false $D^0$ candidates estimated from $D^0$ mass sidebands and
(open points) the mass distribution from
wrong-sign $\overline{D}{}^0 p$ candidates.
}
\label{fig.D0p}
\end{figure}

Using 287\invfb\ of data, \babar\ has investigated~\cite{lambdac2940}
the $D^0p$ final state in \ccbar\ continuum events.
Fig.~\ref{fig.D0p} shows the $D^0p$ invariant mass spectrum. Two prominent structures
are immediately visible: one near a $D^0p$ mass of 2880\mevcc\ 
(${\cal O}(3000)$ events), 
the other near 2940\mevcc\ (${\cal O}(2300)$ events).
No structure is observed in the $D^+p$ final state, 
indicating that there is no doubly-charged (isospin) partner for either state.
We identify the first structure with the known $\Lambda_c(2880)$,
which is observed in this final state for the first time,
and measure its mass and width (preliminary)
\begin{eqnarray}
m      & = & (2881.9 \pm 0.1 \pm 0.5)\mevcc, \nonumber \\
\Gamma & = & (5.8 \pm 1.5 \pm 1.1)\mev\,,
\label{eq.36}
\end{eqnarray}
the precision of which is a large improvement over the PDG~\cite{pdg} values.
In addition, this analysis determines the state to be an isoscalar.

The second structure we identify as a new charmed baryon, which we label
$\Lambda_c(2940)$, with mass and width (preliminary)
\begin{eqnarray}
m      & = & (2939.8 \pm 1.3 \pm 1.0)\mevcc, \nonumber \\
\Gamma & = & (17.5 \pm 5.2 \pm 5.9)\mev\,.
\label{eq.37}
\end{eqnarray}
The placement of this new state in the SU(4) multiplets (including its spin-parity)
is not yet known.

\begin{figure}[h]
\centering
\includegraphics[width=80mm]{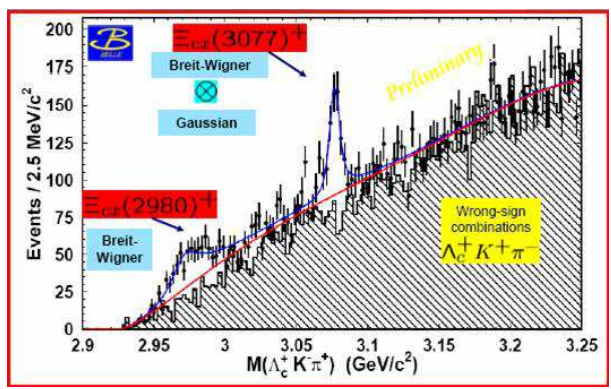}
\caption{$\Lambda_c^+ K^- \pi^+$ invariant mass spectrum.
The shaded area shows (doubly-charged) wrong-sign combinations.
}
\label{fig.lambdacx}
\end{figure}

The Belle collaboration has used 462\invfb\ of \ccbar\ continuum data
to analyze the $\Lambda_c^+ K^- \pi^+$ final state
in search for the doubly-charmed 
$\Xi_{cc}(3520)^+$, 
reported~\cite{selex3520} 
by the SELEX collaboration. 
They find no evidence for this state, but in the process observe 
two new charmed-strange baryons,
$\Xi_{cx}(2980)^+$ and $\Xi_{cx}(3077)^+$, see Fig.~\ref{fig.lambdacx},
where the subscript $x$ denotes the total angular momentum
of the light quark system.
Signal yields for the two states are 
${\cal O}(400)$ and ${\cal O}(300)$ events, respectively, 
and their (preliminary) measured masses and widths are
\begin{eqnarray}
m      & = & (2978.5 \pm 2.1 \pm 2.0)\mevcc, \nonumber \\
\Gamma & = & (43.5 \pm 7.5 \pm 7.0)\mev\,,
\label{eq.39}
\end{eqnarray}
and 
\begin{eqnarray}
m      & = & (3076.7 \pm 0.9 \pm 0.5)\mevcc, \nonumber \\
\Gamma & = & (6.2 \pm 1.2 \pm 0.8)\mev\,.
\label{eq.40}
\end{eqnarray}
Again, it is not yet known how these states fit into the multiplet scheme.

\section{Conclusions} 

Over the past few years, there has been a lot of progress in the areas
of charm and charmonium spectroscopy. 
\babar\ and Belle have observed a number of $X/Y/Z$ states in reactions likely
to produce \ccbar\ states and have measured many of their properties. 
The $X(3940)$ and $Z(3930)$ are likely to be conventional charmonium states. 
The $Y(3940)$ needs confirmation whether it really is a (separate) resonant
state. 
The $X(3872)$ and $Y(4260)$ do not fit into the scheme of conventional 
charmonium states. They have become good candidates for unconventional
explanations, for example, in term of 
$\bar D^0$-$D^{*0}$ molecule and $\ccbar g$ hybrid states.

The properties of the
$D^*_{sJ}(2317)$ 
and 
$D_{sJ}(2460)$ mesons---mass, width, decay modes, and spin-parity---have by now 
been determined quite precisely. 
Apart from the unexpectedly low masses and the smallness of certain
radiative decays, the properties of both states are consistent with
their interpretation as conventional P-wave $c \bar s$ mesons with
$\JP = 0^+$ and $\JP = 1^+$, respectively.

\babar\ and Belle keep discovering new charmed baryons, often with
signals of hundreds or even thousands of events, but the interpretation
of these states in terms of SU(4) multiplets is often uncertain because
no spin-parity information is available.

As the $B$-Factory data samples keep growing, one can expect progress
in understanding the nature of all these states to continue, as well as
more discoveries of states of conventional and unconventional nature.

% If you have acknowledgments, this puts in the proper section head.
\bigskip % extra skip inserted
\begin{acknowledgments}
I would like to thank my colleagues at \babar\ and Belle for helpful
discussions, in particular Bill Dunwoodie, David Williams, and Bostjan Golob.
%Work supported by Department of Energy contract DE-AC02-76SF00515.
\end{acknowledgments}

\bigskip % extra skip inserted
% Create the reference section using BibTeX:
%\bibliography{basename of .bib file}

\end{document}
%
% ****** End of file template.aps ******